\documentclass[journal,usenatbib]{rmaa}

\usepackage{paralist}

\usepackage{psfrag,color}

\usepackage[latin1]{inputenc}
\usepackage[T1]{fontenc}
\usepackage{rotating}

\title{RR Lyrae stars in the globular cluster Palomar 2} 

\author{
  A. Arellano Ferro,\altaffilmark{1}
  I. Bustos Fierro,\altaffilmark{2}
  S. Muneer,\altaffilmark{3}
  S. Giridhar\altaffilmark{3} }

\altaffiltext{1}{Instituto de Astronom\'ia, Universidad Nacional Aut\'onoma de M\'exico, Ciudad Universitaria, C.P. 04510, M\'exico.}

\altaffiltext{2}{Observatorio Astron\'{o}mico, Universidad Nacional de
C\'{o}rdoba,  Laprida~854, X5000BGR, C\'{o}rdoba, Argentina.}

\altaffiltext{3}{Indian Institute of Astrophysics, Bangalore, India.}

\shortauthor{Arellano Ferro et al. }
\shorttitle{RR Lyrae stars in Pal 2}

\abstract{A CCD \emph{VI} imaging  time-series over 11-year is employed to explore the light curves of stars in the field of Palomar 2. We discovered 20 RRab and 1 RRc variables. A revision of $Gaia$-DR3 data enabled us to identify 10 more variables and confirm the RRab nature of 6 of them and one RGB. The cluster membership is discussed and 18 variables are most likely cluster members. The Fourier light curve decomposition for the 11 best quality light curves of cluster member stars leads to independent estimates of the cluster distance 27.2 $\pm$ 1.8 kpc and [Fe/H]$_{\rm ZW}$=-1.39$\pm$ 0.55. \textcolor{blue}{We confirm the cluster as of the Oo I type.}}
\resumen{Empleando una serie temporal de 10 a\~nos de im\'agenes CCD \emph{VI}, exploramos las curvas de luz de estrellas en el campo del c\'umulo. Descubrimos 20 RRab y  1 RRc. Una revisi\'on de los datos de $Gaia$-DR3 permiti\'o identificar 11 variables m\'as y confirmar la naturaleza RRab de 6 de ellas y una RGB. Presentamos un an\'alisis de membres\'ia y concluimos que al menos 18 de estas variables pertenecen al c\'umulo. La descomposici\'on de Fourier de las curvas de luz de mejor calidad de 11 RR Lyrae miembro conduce a estimaciones independientes de la distancia 27.2 $\pm$ 1.8 kpc y metalicidad [Fe/H]$_{\rm ZW}$=-1.39$\pm$ 0.55 medias para el c\'umulo. \textcolor{blue}{Confirmamos que el c\'umulo es del tipo Oo I.}}

\addkeyword{Stars: variables: RR Lyrae}
\addkeyword{Globular Clusters: Individual: Pal 2}

\begin{document}
\maketitle

\section{Introduction}
\label{intro}

The globular cluster Palomar 2 is a distant (~30 kpc) stellar system in the direction of the Galactic anticenter and close to the Galactic plane  ($l = 170.53^{\circ}$, $b = -9.07^{\circ}$). It is buried in dust with $E(B-V) \sim 0.93$ and \textcolor{blue}{shows evidence} of differential reddening \citep{Bonatto2020}. It is therefore a faint cluster with the HB at about $V \sim 21.5$ \citep{Harris1996}. Most likely due its faintness no variables in the cluster have ever been reported. 

In the present paper we take advantage of a 11-year long time-series of CCD \emph{VI} data, analyzed in the standard Differential Imaging Approach (DIA), to explore the light curves of nearly 500 stars in the field of view (FoV) of the cluster. We have found 21 new RR Lyrae stars (V1-V14 and SV1-SV7 in Table \ref{variables}). \textcolor{blue}{In conjunction} with the $Gaia$-DR3 variabilty index, we confirm the RRab nature of 6 more stars (G3, G11, G12, G13, G18 and G23), plus 1 RGB (G17), for a total 28 variables in the field of view of our images. In what follows, we argue in favour of the membership of 18 of them and \textcolor{blue}{present} their light \textcolor{blue}{curve} and ephemerides. The mean distance and [Fe/H] of the cluster shall be calculated by the Fourier decomposition of RRab stars with the best quality light curves.

\section{Observations and Data Reductions}
\label{observations}

The data were obtained between December 12, 2010 and February 12, 2021 with the 2.0-m telescope at the Indian Astronomical Observatory (IAO), Hanle, India. The detector used was a SITe ST-002 2Kx4K with a scale of
0.296 arcsec/pix, for a field of view of approximately 10.1$\times$10.1 arcmin$^2$.
From October 14, 2018 and February 17, 2020 the detector used was a Thompson
grade 0 E2V CCD44-82-0-E93 2Kx4K with a scale of 0.296 arcsec/pix, or a FoV of approximately 10.1$\times$10.1~arcmin$^2$.
A total of 197 and 240 images were obtained in $V$ and $I$ filters, respectively.

\subsection{Difference imaging analysis}

The image reductions were performed employing the software Difference Imaging Analysis (DIA) with its pipeline implementation DanDIA (\citealt{Bramich2008}; \citealt{Bramich2013,Bramich2015}) to obtain high-precision photometry of all the point sources in the field of view (FoV) of our CCD. This allowed us to construct an instrumental light curve for each star. For a detailed explanation \textcolor{blue}{of} the use of this technique, the reader is referred to the work by \citet{Bramich2011}. 




\subsection{Transformation to the standard system}
Since two different detectors were used to achieve the observations as described in the previous section, we treated the transformation to the standard system as two independent instruments. Otherwise, the procedure was the standard one described into detail in previous publications, in summary; we used local standard stars taken from the catalog of Photometric Standard Fields \citep{Stetson2000} to set our photometry into the $\emph{VI}$ Johnson-Kron-Cousin standard photometric system \citep{Landolt1992}.

The transformation equations carry a small but mildly significant colour term and are of the form:  $V-v = A (v-i)+B$ and $I-i = C (v-i)+D$  for each filter respectively. The interested reader can find the details of this transformation approach in \citet{Yepez2022}.

\section{Star membership using $Gaia$-eDR3}
\label{gaia}

We have made use of the latest data release $Gaia$-DR3 \citep{Gaia_edr32021} to \textcolor{blue}{perform} a membership analysis of the stars in the field of Pal 2. To this end, we employed the method of \citet{Bustos2019}, which is based on
the Balanced Iterative Reducing and Clustering using Hierarchies (BIRCH) algorithm developed by \citet{Zhang1996}. The method and our approach to it have been described in a recent paper by \citet{Deras2022}. We recall here that our method is based on a clustering algorithm at a first stage and a detailed analysis of the residual overdensity at a second stage; member stars extracted in the first stage are labeled M1, and those extracted in the second stage are labeled M2. Stars without proper motions were retained labeled as unknown membership status or UN.

The analysis was carried out for a 10 arcmin radius field centered in the cluster. We considered 1806 stars with measured proper motions \textcolor{blue}{of} which 407 were found to be likely members. \textcolor{blue}{Out} of these only 288 were in the FoV of our images, for which we could produce light curves.

From the distribution of \textcolor{blue}{the} field stars in the phase space we estimated the number that is expected to be located in the same region of the sky and of the VPD as the extracted members, therefore they could have been erroneously labelled as members. Within the M1 stars the resulting expected contamination is 36 (11\%) and within the M2 stars it is 87 (7\%); therefore, for a given extracted star its probability of being a cluster member is 89\% if it is labeled M1, or 93\% if it is labeled M2.

\begin{figure*}
\begin{center}
\includegraphics[width=\textwidth, height=8cm]{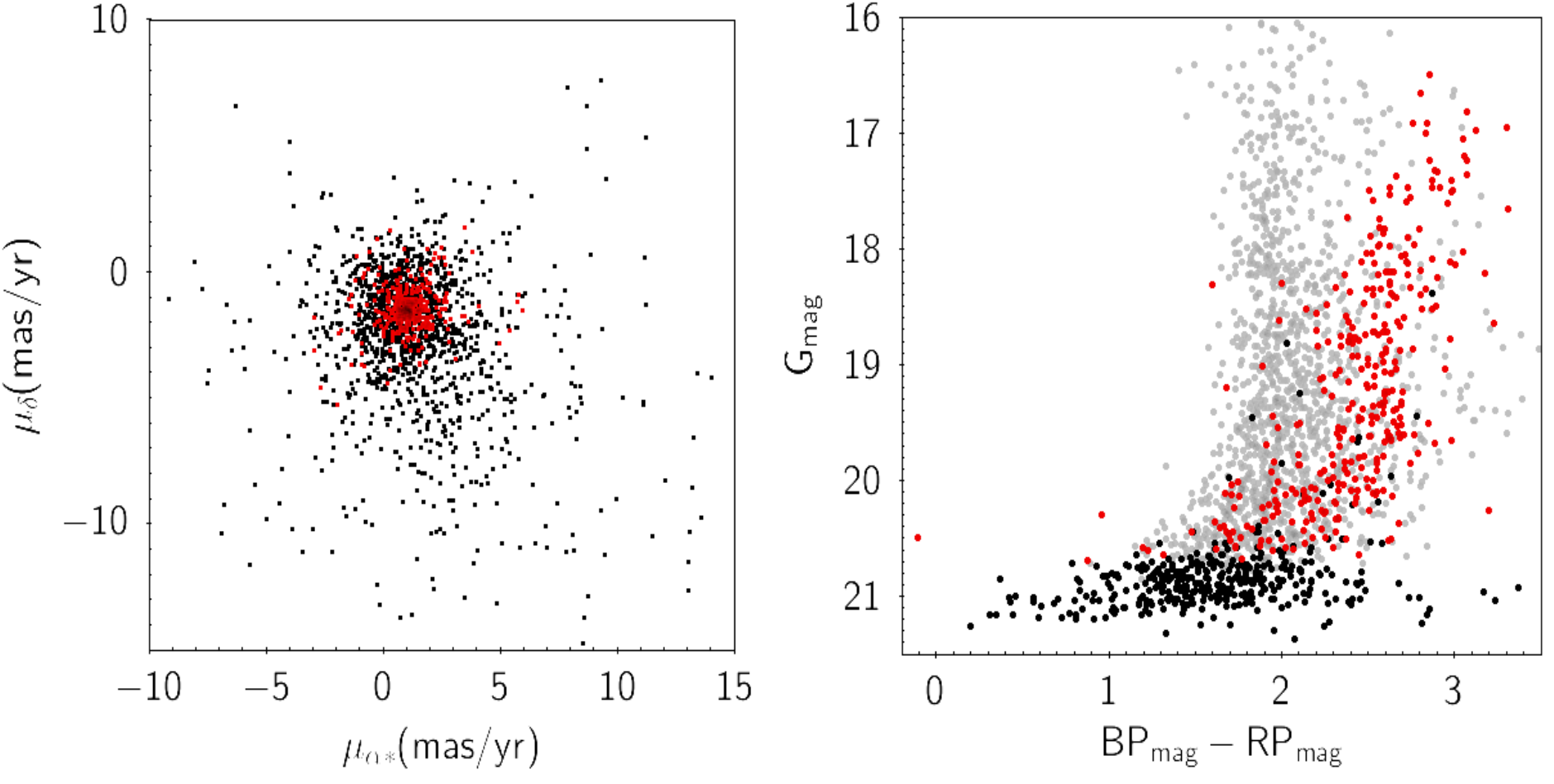}
\caption{VPD and Gaia CMD of Pal 2. In the left panel, black and red points represent field and member stars respectively, as extracted by our analysis of  the $Gaia$DR3 proper motions. In the right panel gray and red symbols are for non member and member stars, while black dots are used of stars without proper motion information, hence their membership status cannot be assigned (UN). }
\label{knownvar}
\end{center}
\end{figure*}

\section{Differential Reddening and the CMD}
\label{CMD}

\begin{figure*}
\begin{center}
\includegraphics[width=\textwidth, height=8cm]{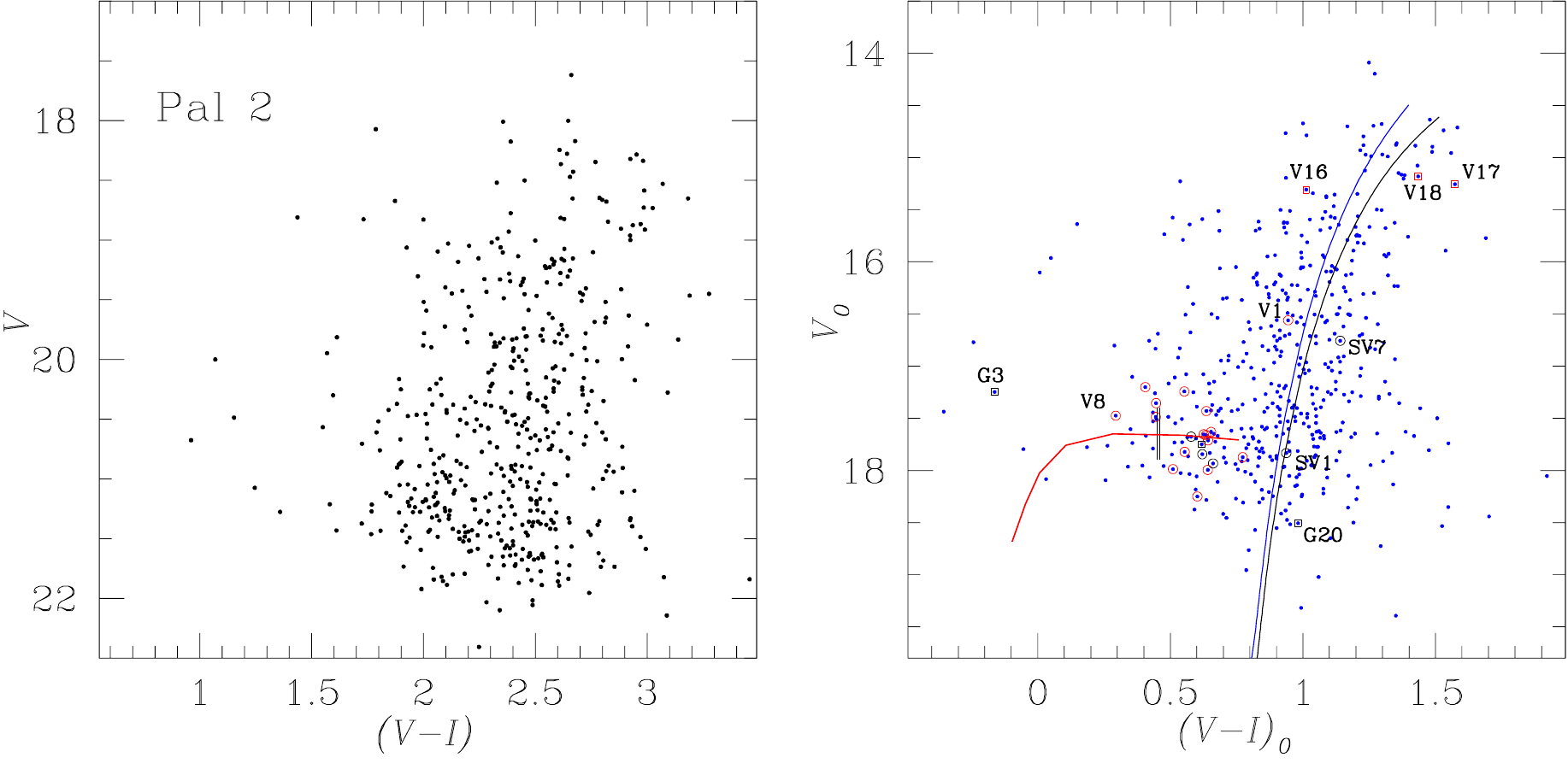}
\caption{The Colour-Magnitude Diagram of Pal 2. The left panel shows all the stars measured in the FoV of our images and illustrates the magnitude and
colour ranges of our data. The right panel has been deferentially dereddened by adopting the reddening map of \citet{Bonatto2020}. We adopted from these authors an average foreground reddening of $E(B-V) = 0.93$. As a reference we included two isochrones from the models of \citet{Vandenberg2014} for [Fe/H]=-1.6 and -2.0 and a theoretical horizontal branch built by \citet{Yepez2022}. Isochrones and HB have been placed at a distance of 26.1 kpc \citep{Bonatto2020}. Variable stars are indicated and discussed in $\S$ \ref{var_star}.}
\label{CMD_Pal2}
\end{center}
\end{figure*}

Palomar 2 is a heavily reddened cluster subject to substantial differential reddening as it is evident in the crowded and deep HST Color Magnitude Diagram (CDM) shown by \citet{Sarajedini2007}. A thorough treatment of the differential reddening in the cluster enabled \citet{Bonatto2020} to produce a reddening map which these authors have kindly made available to us. In Figure \ref{CMD_Pal2} the observed CMD and the dereddened versions are shown. To deredden the CMD, the differential reddening map was added to a forground reddening of $E(B-V)=0.93$.

\begin{table*}
\scriptsize
\begin{center}
\caption{Data of variable stars in the FoV of our Pal 2 images.}
\label{variables}
\begin{turn}{90}
\begin{tabular}{cccccccccccc}
 \hline
ID& Gaia  & Type & P &E$_0$&$V$ & $V$ Amp& RA & DEC & $P_{\rm Gaia}$ &  Membership& Gaia \\
 &variable  &  &  (d)& (+2450000)& (mag)&(mag)& (J2000.0)&(J2000.0)& (d)& status& number \\  
 \hline
V1& & RRab& 0.542848 &6312.3363&20.534&0.805&  4:46:03.57 &+31:22:45.8& & M1 &159504640014524672\\
V2& G5& RRab& 0.551396& 5542.2114 &21.342 &1.056 & 4:46:04.60& +31:23:41.5& 0.5513624 & M1& 159504747388302336\\
V3& & RRab& 0.554363 &6948.4976&21.792&0.951&4:46:05.53 &+31:23:29.0&  & M1& 159504747388520064\\
V4& G14& RRab& 0.651889 &5912.2228&21.413 &0.814 &  4:46:05.61 &+31:23:43.2 &0.6518656 & M1& 159504747387726464\\
V5 & G4 &RRab& 0.511639 &8896.2470&21.382&0.997 &4:46:07.02& +31:23:13.5 &0.5067667 & M2& 159504678667943552\\
V6 & G21&RRab& 0.553259& 9258.3356&21.461&1.168 &4:46:07.82& +31:23:07.7& 0.5532034 & M2& 159504678668831872\\
V7 & G16&RRab& 0.655812& 8407.3827&20.925&0.914 &4:46:08.11 &+31:23:37.1 &-- &  M1& 159504678667937024\\
V8 & G7&RRc&  0.373408 &5542.2114&20.757 &0.548 &4:46:08.06 &+31:22:21.7& --&   M1& 159501689370744192\\
V9 & G6 &RRab& 0.629619 &8896.1493&21.521&0.787 &4:46:08.24 &+31:23:09.3& 0.6129630 & M1& 159504674373384320\\
V10& G8&RRab& 0.685890&5912.3072&20.700 &0.512 &4:46:09.11& +31:22:38.0 &0.6858277&  M1& 159501723731340288\\
V11& G19&RRab& 0.575280& 6222.3870&20.673 &0.842 &  4:46:10.58 &+31:22:35.0& 0.5752915 & M1&159501719435472896\\
V12 &G9&RRab&  0.583630& 6633.3246&20.894 &0.603 &   4:46:12.82 &+31:22:26.3& 0.5953860 & M1& 159501650715992064\\
V13 &  & RRab& 0.546972& 6948.4441&21.327 &0.887 &4:46:07.17 &+31:23:15.5& &  M2& 159504678668829184\\
V14 &G1 &RRab& 0.574697 &6948.4591&21.842 &1.610 &4:46:07.21 &+31:22:47.2 &0.5513435 & M2 & 159504678667961856\\
V15&G12 &RRab&   0.508471& 8781.4301 &20.918 &0.323 &  4:46:05.00 &+31:22:52.9 &--  & M1& 159504644308236672\\
V16 &G13 &RR?&0.490213  &5912.1144&19.179 &0.330 &4:46:04.64 &+31:22:42.0& -- &  M1& 159504644308250624\\
V17 & G17&RGB& & &19.0 &0.9 &4:46:02.96&  +31:23:09.2& -- &  M1 &159504708733123200\\
V18 & G11 &RR?& 0.510211 &5912.1144&18.876 &0.768 &4:46:05.85 &+31:23:03.3& --&  M1& 159504644308215808\\
SV1 &&RRab& 0.588566& 6634.1554&21.267 &1.024 &  4:46:04.22 &+31:22:34.8& & UN &159504644309111808\\
SV2 &&RRab &0.537325& 8406.4629&21.876 &1.299 &  4:46:06.39 &+31:23:54.0 & &  UN &159504747388298112\\
SV3& & RRab & 0.661914& 5868.4136&21.517 &1.077 &   4:46:03.96 &+31:23:16.2& &  FS&159504713028573696\\
SV4& & RRab  & 0.587210 &8407.3175&21.585&1.363 &  4:46:06.56 &+31:23:27.2 &  &FS&  159504674373556992\\
SV5& G15&RRab  & 0.490941 &6221.4206&21.312&1.391 &  4:46:09.04 &+31:23:12.8 &0.4909349&  FS& 159504678668828160\\
SV6& G10&RRab  & 0.570669 &6946.4683 &20.840&0.960 &  4:46:12.31& +31:22:45.3 &0.5706582&  FS& 159501723731332480\\
SV7& &RRab&0.551215& 6634.1714&19.274&1.371 &4:46:13.65 &+31:24:11.5&   &   FS& 159506190497880832\\
&G3 & RRab&   0.531512 &6633.3479&20.486 &0.769&  4:45:57.72 &+31:24:19.0& 0.5242873&  FS& 159504987906469248\\
&G18& RRab&   0.562320& 6223.3662&20.700 &0.576&4:46:09.50 &+31:23:01.9& 0.5623196&FS &159501723732926208 \\
&G23&RRab&    0.595453& 6633.3810&21.178 &1.104 &  4:45:59.23 &+31:22:53.4& 0.56065912& FS& 159504609949939072 \\
&G2$^1$ &&&&&& & & &&159501655012584064\\
&G20$^2$& & &  &21.513 &  &4:45:56.77& +31:21:09.0&0.59148323& FS &159504128913233536\\
&G22$^1$ &&&&&&&\\   

 \hline
\hline
\end{tabular}
\end{turn}

\end{center}
\raggedright
\center{
1. Out of our FoV. 2. Not measured by our photometry.
}

\end{table*}

\section{The variable stars in Pal 2}
\label{var_star}

No variable stars in Pal 2 have been thus far reported. The case of Pal 2 is a particularly challenging one since
the cluster is not only distant but it is also behind a heavy dust curtain, its horizontal branch (HB) is located below 21 mag. We have occasionally taken CCD \emph{VI} images of Pal 2 since 2010 and until 2021 and we have attempted to take advantage of this image collection to search for variables in the FoV of the cluster. We were able to measure 400-500 point sources in the $V$ and $I$ images that span a range in magnitude and colour shown in the left panel of  Fig. \ref{CMD_Pal2}. The HB being located at the bottom of the stellar distribution, we are in fact working at the very limit of our photometry in order to detect cluster member RR Lyrae.   

To search for variability we proceeded as follows. 
By using the string-length method (\citealt{Burke1970}, \citealt{Dworetsky1983}), we phased each light curve in our data with a period varying between 0.2 d and 1.0 d, a range adequate for RR Lyrae stars, in steps of $10^{-6}$ d. For each phased light curve, the length of the line joining consecutive points, called the string-length, represented by the parameter $S_{Q}$ was calculated. The best phasing occurs when $S_{Q}$ is minimum, and corresponds to the best period that our data can provide. A detail visual inspection of the best phased light curve helped \textcolor{blue}{ to confirm} the variability of some stars.
We noticed however, that the seasonal scatter of the  light curve could vary largely depending mainly of the prevailing seeing conditions and crowdedness of a particular star, a situation that worsens near the core of the cluster. Therefore, it may happen that in some seasons the light curve variation is dubious but extremely clear in the runs of best quality, which turned out to be from the 2013 and 2018-2020 seasons.  

\textcolor{blue}{With} the above method we discovered 21 RR Lyrae variables, mostly of the RRab type. Confronting with the membership analysis described in $\S$ \ref{gaia}, we concluded that 14 of them were likely cluster members.
The latest $Gaia$-DR3 enable us to search for stellar variability flags in the field of Pal 2. In fact Gaia flags 22 variables. A cross-match with our variables list show 12 matches; we found some variables not marked by $Gaia$ and {\it a posteriori} we confirmed the variability of a few $Gaia$ sources not previously detected by us. 

\begin{figure*}
\begin{center}
\includegraphics[width=\textwidth, height=18cm]{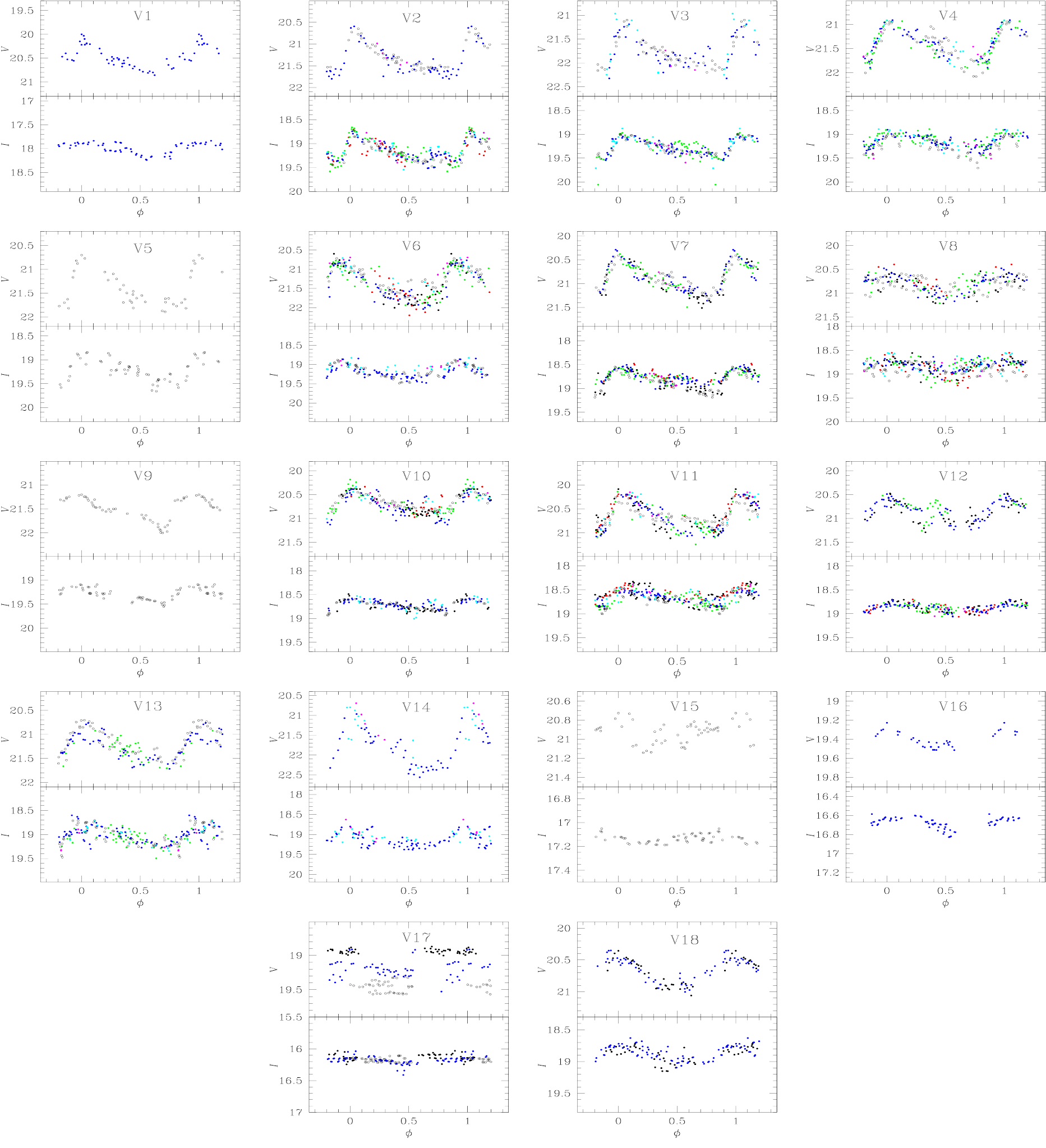}
\caption{Light curves of cluster member variables in Pal 2. Different colors are used for yearly seasons. From the plots it is obvious that the best quality data are from the 2013 (blue dots) and 2018-2020 (open circles) seasons. The rest, although more scattered, do follow and confirm the variations.}
\label{membervars}
\end{center}
\end{figure*}

In Table \ref{variables} we list the 32 variables in the field of Pal 2. The table is organized as follows.
We have given the name with a prefix "V" only to those stars that seem likely cluster members (status M1 or M2), 18 in total, V1-V18.
Arbitrarily, we identified the $Gaia$ variables as G1-G22. This identification is listed in column 2.
In the bottom 14 rows of Table \ref{variables} we list the likely non-members (status FS). For non-member variables detected by us, we used the nomenclature with the prefix "VS".

\subsection{Variables in the CMD}

In the right panel CMD of Fig. \ref{CMD_Pal2} all variable stars have been circled with a red circle if cluster member or a black circle otherwise. As a reference we included two isochrones from the models of VandenBerg et al. (2014)
for [Fe/H]=$-1.6$ and $-2.0$ and a theoretical horizontal branch built by Yepez et al. (2022). Isochrones and HB were
placed at a distance of 26.1 kpc \citet{Bonatto2020}. It is heartening to see nearly all the RR Lyrae stars fall in the whereabouts of the HB.  
In the following section we address some peculiar individual cases.

\begin{figure*}
\begin{center}
\includegraphics[width=\textwidth, height=10cm]{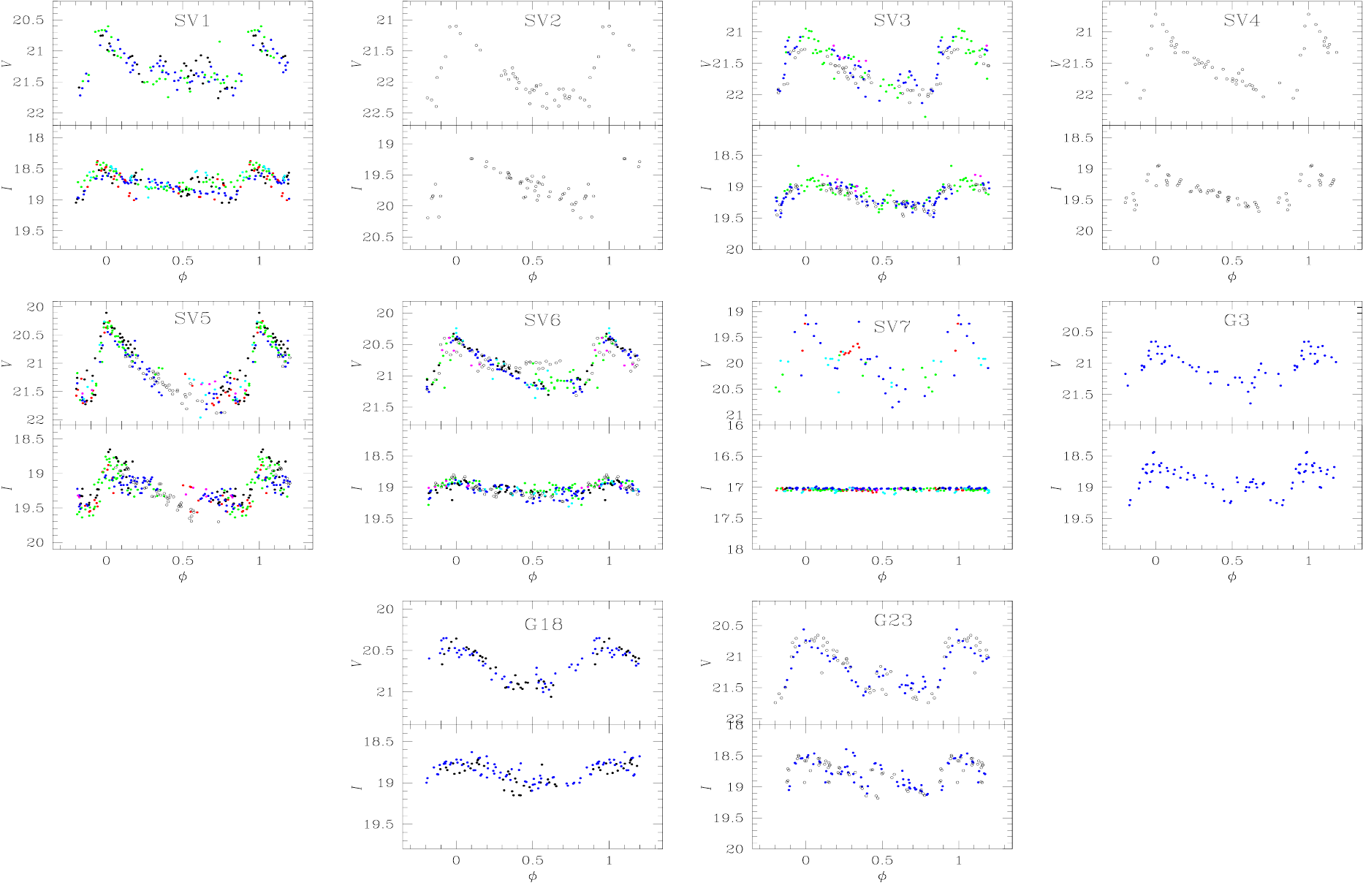}
\caption{Light curves of variables in the field of Pal 2. They are most likely field stars, see $\S$ \ref{var_star} for a discussion. The colour code is as in Fig. \ref{membervars}.}
\label{NonMember}
\end{center}
\end{figure*}

\subsection{Individual cases}
V1. Its position on the CMD above the HB and in the mid-RGB is intriguing since the light curve and period suggest this star to be a member RRab star. An alternative possibility is that the star is a binary. Our data are not sufficient to explore this possibility.

V16, V17 and V18. Their position on the CMD near the tip of the RGB suggests these stars to be red giant variables. However our photometry was not intensive enough to confirm a long term variability. Alternatively, we were able to identify short therm variations in V16 and V18 (see Fig. \ref{membervars}). V17 light curve is in fact consistent with that of a long-term RGB. 

SV1. It is a clear RRab star falling too much to the red of the HB. The star is not a cluster mamber.

SV7. We have detected clear RRab-like variations in our $V$ data. However no variation is seen in the $I$ data. While variations might be spurious, we retain the star as a candidate variable to be confirmed.

SV4, SV5, SV6 and G23. These are the four non-member stars hence identified by black circle or square in the DCM. However they lie very near the HB.
Their non membership status was assigned by the statistical approach to their proper motions but they might be cluster members.

G3 and G20. G3 is a clear RRab star not a cluster member. For G20 we got a very noisy light curve that that makes its classification very difficult, however, the star is likely a non-member.

\section{{The Oosterhoff type of Pal 2}}
\label{Oosterhoff}
\begin{figure}
\begin{center}
\includegraphics[width=8cm, height=12cm]{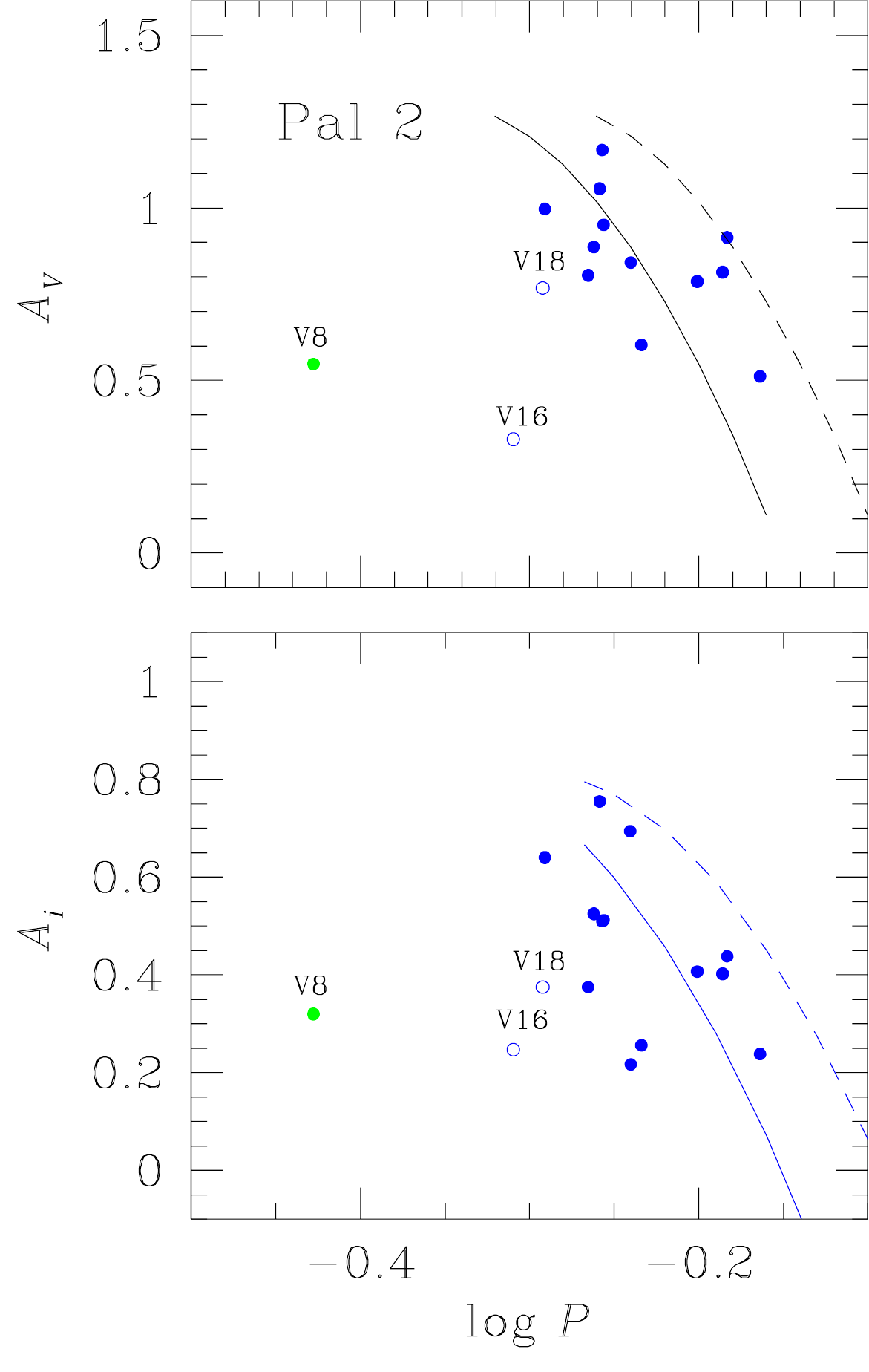}
\caption{\textcolor{blue}{Distribution of RR Lyrae stars in the amplitude-period diagram. The solid sequences correspond to the unevolved stars typical of Oo I type clusters \citep{Cacciari2005}. The upper sequence corresponds to evolved stars of the Oo II clusters \citep{Kunder2013}. V8 is a RRc star. See $\S$ \ref{Oosterhoff} for a discussion on V16 and V18.}}
\label{bailey}
\end{center}
\end{figure}

\textcolor{blue}{The average period of the member RRab star listed in Table \ref{variables} is 0.55 days which indicates that Pal 2 is of the Oo I type. We can further confirm this from the distributions of the RRab stars in the Amplitude-Period or Bailey diagram, shown in Fig. \ref{bailey}. Given the dispersion of the light curves, the amplitude distribution is also scattered, however it is clear that the RRab stars follow the expected sequence for unevolved stars typical of OoI clusters \citep{Cacciari2005}, in both $V$ and $I$ bands . The upper sequence corresponds to evolved stars of the Oo II clusters \citep{Kunder2013}. Hence, Pal 2 is a Oo I type cluster.
We note the stars V16 and V18, whose nature is not clear due to their position in the RGB and short period ($\S$ \ref{var_star}), do not follow the general trend rather confirming that they are not field RR Lyrae stars. Alternatively they may be binary stars. Further observations may be required to provide a proper classification.}

\section{Cluster distance and metallicity from member RR Lyrae stars}\label{fourier}

Although the scatter of all \textcolor{blue}{these} faint cluster member stars may be large, we attempted an estimation of the mean distance and [Fe/H] via the Fourier light curve decomposition. This approach has been amply described in previous papers. Both the method details and the specific calibrations for $M_V$ and [Fe/H] for RRab stars can be found in a recent paper by  \citet{Arellano2022}.

We selected the RRab members with best quality light curves and restricted the Fourier approach to this sample. These are the variables V2-V13 shown in Fig. \ref{membervars}. The mean values for the distance modulus  $(V-M_V)_o$=17.18 and [Fe/H]$_{\rm ZW}$= $-1.39 \pm 0.55$ were found. The corresponding distance is $27.2 \pm 1.8$ kpc for a foreground reddening of $E(B-V)=0.93$ plus the differential values according to the reddening map of \citet{Bonatto2020}. The quoted errors are the standard deviation of the mean, they are a bit too large but given the faintness of the stars and their consequent photometric scatter, the results are remarkably in good agreement with independent determinations: $(V-M_V)_o$=$17.1 \pm 0.1$ and [Fe/H]=-1.3 \citep{Harris1997}; Fe/H]$_{\rm ZW}$=$-1.68\pm 0.04$ \citep{Ferraro1999}; or $d=27.2$ kpc and [Fe/H]=$-1.42$ listed by \citet{Harris1996} (2010 update).

\begin{figure*}
\begin{center}
\includegraphics[width=\textwidth, height=9cm]{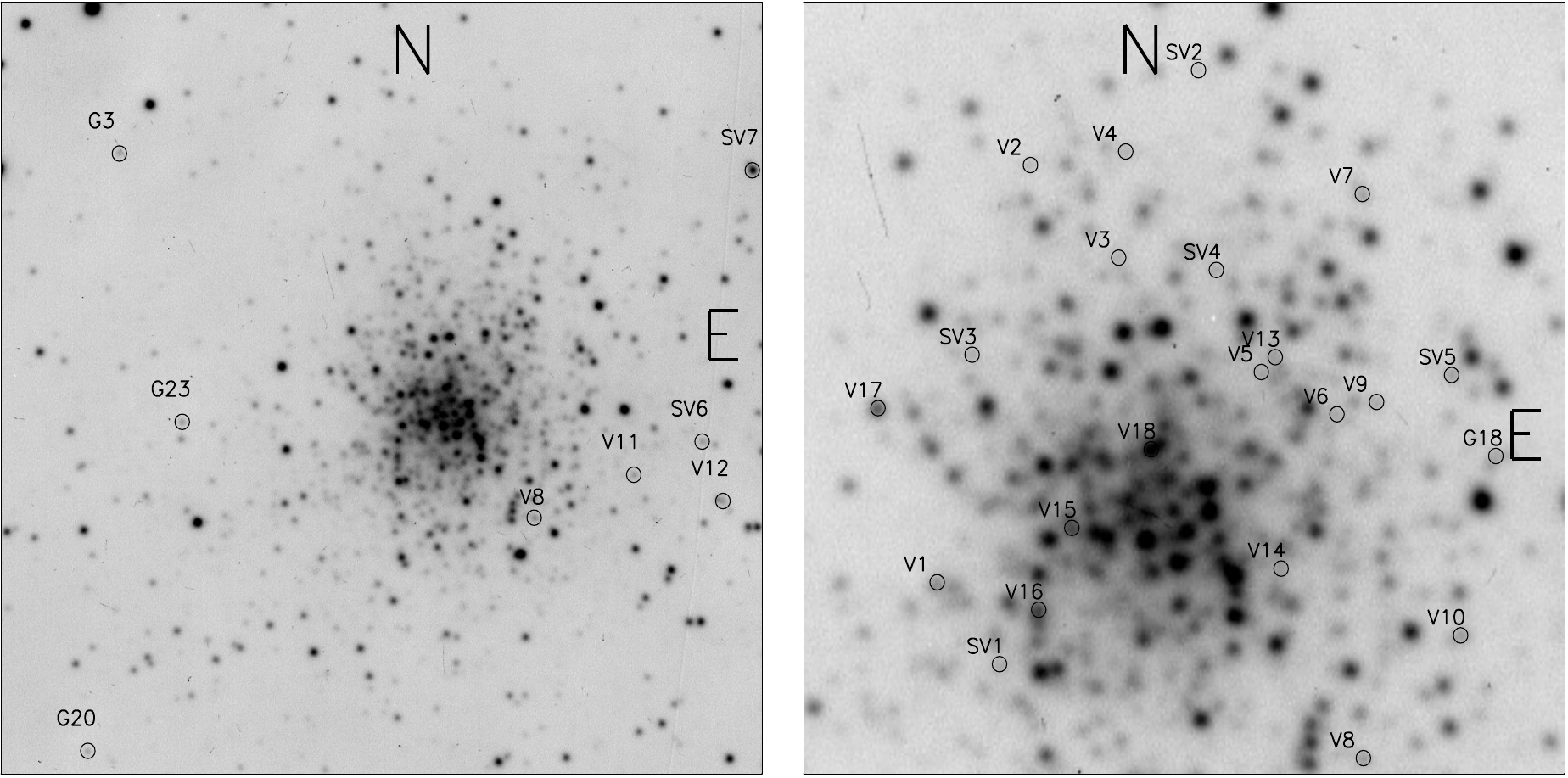}
\caption{Identification chart of variables in our FoV of Pal 2. The left panel shows a field of  $4.1 \times 4.1$ arcmin$^2$. The right panel is about $1.7 \times 1.7$ arcmin$^2$ . Expansion of the digital version is recommended for clearness.}
\label{idchart}
\end{center}
\end{figure*}

\section{summary}

We have found and identified 32 variables in the field of the globular cluster Palomar 2. A membership analysis based on $Gaia$-DR3 proper motions and the positioning of the variables in the corresponding intrinsic CMD, demonstrates that at least 18 of these variables are cluster members. Most of the detected variables are of the RRab type but one RRc and at least one RGB were identified.

The mean cluster distance and metallicity, estimated from the Fourier light curve decomposition of 11 cluster member RRab stars with the best quality available data, lead to $d= 27.2 \pm 1.8$ kpc and metallicity $-1.39 \pm 0.55$ in reasonable agreement with the previous estimates. A detailed finding chart of all these variables is provided.

\label{summary}

\section{ACKNOWLEDGMENTS}
This project was partially supported by DGAPA-UNAM (Mexico) via grants IG100620. AAF is thankful to Mr. G.A. Garc\'ia P\'erez and Mr. G. R\'ios Segura for computational help. The
facilities at IAO and CREST are operated by the Indian Institute
of Astrophysics, Bangalore, we are grateful for the observing time allocated and for the valuable help of the support staff.

\bibliographystyle{rmaa}
\bibliography{Pal2}


\end{document}